\documentclass[fleqn,11pt]{article}

\usepackage{amsmath,amssymb}
\usepackage{graphicx}
\usepackage[top=0.75in, bottom=0.75in, left=0.75in, right=0.75in, dvips]{geometry}
\usepackage{color}

\begin{document}

\title{{\sf Effect of Multiple Higgs Fields on the Phase Structure of the SU(2)-Higgs Model}}

\author{
Mark Wurtz\thanks{Department of Physics and
Engineering Physics, University of Saskatchewan, Saskatoon, SK,
S7N 5E2, Canada},
Randy Lewis\thanks{Department of Physics and Astronomy, York University, Toronto, ON, M3J 1P3, Canada},
T.G.\ Steele\thanks{Department of Physics and
Engineering Physics, University of Saskatchewan, Saskatoon, SK,
S7N 5E2, Canada} }

\maketitle

\begin{abstract}
The $SU(2)$-Higgs model, with a single Higgs field in the fundamental representation and a quartic self-interaction, has a  Higgs region  and a confinement region which are analytically connected in the parameter space of the theory; these regions thus represent a single phase.  The effect of multiple Higgs fields
on this phase structure is examined via Monte Carlo lattice simulations. 
For the case of $N\ge 2$ identical Higgs fields, there is no remaining analytic connection between the Higgs and confinement regions, at least when Lagrangian terms that directly couple different Higgs flavours are omitted. 
An explanation of this result in terms of enhancement from overlapping phase transitions
is explored for $N=2$ by introducing an asymmetry in the hopping parameters of the Higgs fields.  It is found that  an enhancement of the phase transitions can still occur for a moderate (10\%)  asymmetry in the resulting hopping parameters.
\end{abstract}

The phase structure of the basic SU(2)-Higgs model ({\it i.e.\ }an $SU(2)$ gauge theory coupled to a single scalar field with a quartic self-interaction and a quadratic term) is clearly of direct relevance to the Higgs sector of the Standard Model, where discussions of spontaneous symmetry breaking and the transition from a symmetric phase to a Higgs phase are paramount.  The addition of extra Higgs fields occurs in a variety of extensions to the Standard Model, including the minimal supersymmetric Standard Model (see {\it e.g.\ }Ref.~\cite{gunion}).

In the case of a single Higgs, the
$SU(2)$-Higgs model has three parameters in its continuum formulation (Higgs self-coupling, gauge coupling, and Higgs quadratic term); in the lattice formulation of these models the three parameters become the Higgs self-coupling $\lambda$, the gauge coupling $\beta$, and the hopping parameter $\kappa$.  For a single Higgs field in the fundamental representation, there exist  theoretical arguments that 
there must be an analytic connection in ($\beta$, $\kappa$, $\lambda$) parameter space
between the Higgs and confinement regions of the theory \cite{fradkin}.
This behaviour is manifested in  lattice simulations
as a phase-transition line that begins at $(\beta=\infty,\kappa_\infty>0)$ and terminates at a point $(\beta\ge 0,\kappa>\kappa_\infty)$, with a general progression of this termination point to smaller values of
$(\beta,\kappa)$ with decreasing $\lambda$ \cite{lang,kuhnelt}.  
The resulting analytic connection between the Higgs and confinement regions (located in the corner of parameter space toward larger $\lambda$, smaller $\beta$, and larger $\kappa$) is consistent with Elitzur's theorem \cite{elitzur} which says that a local gauge symmetry cannot break spontaneously. 
 For a recent discussion of the breaking of global subgroups within the $SU(2)$-Higgs model and their 
connection to the phase diagram, see 
Ref.\ \cite{greensite}.  We note in passing that Ref.\ \cite{greensite} also mentions the $SU(2)$-Higgs model where the Higgs field is in the adjoint representation rather than the fundamental; in that case the theoretical arguments of \cite{fradkin} do not apply and the center symmetry  can be broken spontaneously.

Many basic properties of the phase structure for a single Higgs field in the fundamental representation are well understood, though there is no consensus on a detailed understanding of the nature of the phase transition (PT) across the entire parameter space.  For small $\lambda$, the PT is demonstrably of first order \cite{jersak} but the PT strength weakens with increasing $\lambda$, complicating the classification of the PT.  The most effective approaches used to address this issue are searches for bimodal (``two-peaked'') distributions and lattice scaling dependence on the  resulting energy gap \cite{languth,languth2,bock,campos} or scaling effects within specific heats (susceptibilities) \cite{languth2,bock,campos,tomiya,bonati}.  For example, Ref.~\cite{bock} concludes that for $\lambda\sim 1$ the PT is first-order for $\beta$ at and slightly above the terminal point of the phase line, but is unable to ascertain the existence of a tricritical point 
where the order of the PT would change; it would be called a critical line in $(\beta,\kappa,\lambda)$ space.
Since the PT decreases in strength with increasing $\lambda$, the $\lambda=\infty$ case presents the greatest challenge in the classification of the PT.  Early work \cite{languth} suggested a weak first-order PT, but a recent study with large lattices presents evidence for a smooth crossover \cite{bonati}.  Furthermore, the $\lambda=\infty$ model  augmented with an additional interaction  leads to a line of first-order PTs which decrease in strength as the additional coupling approaches zero \cite{campos}.

In this paper we study the phase structure of the $SU(2)$-Higgs model with multiple Higgs fields in the fundamental representation. For simplicity, we will omit Lagrangian terms which would mix more than one flavour of Higgs.  We find that the phase transition line extends all the way to $\beta=0$ from $\beta=\infty$ for two or more Higgs fields regardless of the numerical value of $\lambda$, in marked contrast to the case of a single Higgs field.  We show that this enhancement of the PT is associated with overlapping PTs, as has also been reported for the multi-Higgs three-dimensional $U(1)$ theory \cite{ono}.\footnote{By overlapping PTs we mean the influence of other Higgs fields on the PT of any one Higgs field.}

The continuum $SU(2)$-Higgs action in Euclideanized space-time is
\begin{equation}
S_c = \int d^4x \, \tfrac{1}{2} Tr \left\{ F^{\mu\nu}F_{\mu\nu} + D_\mu\phi^{c\dag}\,D^\mu\phi^c + \mu_0^2\phi^{c\dag}\phi^c + \lambda_0(\phi^{c\dag}\phi^c)^2  \right\}~,
\end{equation}
where the Higgs field is in the fundamental representation of the gauge group.
Including multiple Higgs fields in which the Higgs particles are not directly coupled to one another, 
\textit{i.e.\ }they interact with each other only through the gauge field, gives us an action of the form,
\begin{equation}
S_c = \int d^4x \, \tfrac{1}{2} Tr \left\{ F^{\mu\nu}F_{\mu\nu} + \sum_{n=1}^{N} \left[ D_\mu\phi_n^{c\dag}\,D^\mu\phi_n^c + \mu_{0,n}^2\phi_n^{c\dag}\phi_n^c + \lambda_{0,n}(\phi_n^{c\dag}\phi_n^c)^2  \right] \right\}  \,
\label{cont_S}.
\end{equation}
In principle, additional quartic interaction terms are possible, including terms which couple different Higgs fields.  
In our exploratory study, we will not consider this large phase space in its entirety.  Instead, we stay within the phase space of Eq.~\eqref{cont_S}.  This is sufficient to observe a qualitative distinction between the single-Higgs and multi-Higgs theories,  
and is also sufficient for our discussion of overlapping PTs.  The extension to a more complete phase space 
({\it e.g.\ }inspired by the renormalizable scalar sector of a non-minimal standard model), beyond Eq.~\eqref{cont_S}, is left for future work.

The discretized version of Eq.~\eqref{cont_S}  for  4-dimensional hypercube with lattice spacing $a$ (and periodic boundary conditions)
is
\begin{equation}
S = \sum_x \left\{ \sum_{\mu > \nu} \beta \left[ 1 - \tfrac{1}{2} Tr \left(  U_{x\mu} U_{x+\hat{\mu}\nu} U_{x+\hat{\nu}\mu}^\dag U_{x\nu}^\dag \right) \right] + \sum_{n=1}^{N} \left[ \lambda (\rho_{x,n}^2 - 1)^2  +  \rho_{x,n}^2  -  \kappa_n \sum_\mu Tr (\phi_{x,n}^\dag U_{x\mu} \phi_{x+\hat{\mu},n})  \right] \right\}  \, .
\label{lattice_S}
\end{equation}
For simplicity,  we have chosen the couplings $\lambda_{0,n}$ so that the discretized theory contains a single four-point coupling $\lambda$
and hence
the hopping parameters $\kappa_n$ are defined by
\begin{equation}
\mu_{0,n}^2 = \frac{1-2\lambda-8\kappa_n}{\kappa_n a^2} \, .
\end{equation}
The
Higgs field has been  expressed in terms of the Higgs length $\rho_x$ and an ``angular component''
$\alpha_x$ which is in the fundamental representation of $SU(2)$:
\begin{equation}
\phi_x = \rho_x\alpha_x \quad , \quad \rho_x>0 \quad , \quad \alpha_x \in SU(2) \, .
\end{equation}
The gauge-invariant link is defined as the gauge-Higgs coupling term with the Higgs length removed
\begin{equation}
L_{x,\mu,n} = \tfrac{1}{2} Tr(\alpha_{x,n}^\dag U_{x\mu} \alpha_{x+\hat{\mu},n})  \, ,
\end{equation}
which, as discussed below, is a very useful quantity for analyzing the state of the Higgs field.

The Monte Carlo calculations we will use to investigate the nature of the Higgs phase transition employ a combination of heatbath steps \cite{heatbath,bunk,fodor} and overrelaxation steps \cite{over_relax} to optimize the updating of the gauge and Higgs fields.  A single Monte Carlo update for the entire system is comprised of one heatbath and one overrelaxation update for both the gauge and Higgs fields.  The heatbath update for the Higgs field is implemented by generating the four real components of $\phi$ according to the distribution
\begin{equation}
dp(\phi) \, \sim \, d^4\phi \exp \left(- \lambda (\rho^2 - 1)^2 - \sum_{m=1}^4 (\phi_m - V_m)^2  \right)  \, , \label{dp_phi}
\end{equation}
where $\phi = \phi_m \tau_m$, $\tau_m = (I,i\vec{\sigma})$, and $V_m$ is from the nearest neighbour interactions,
\begin{equation}
 V_m = \frac{\kappa}{2} \, Tr \sum_{\mu=1}^4 \tau_m (\phi_{x+\hat{\mu}}^\dag U_{x\mu}^\dag + \phi_{x-\hat{\mu}}^\dag U_{x-\hat{\mu}\mu} )  \, . 
\end{equation}
The distribution in Eq.~\eqref{dp_phi} may be rearranged so that it is expressed in the form
\begin{equation}
dp(\phi) \, \sim \, d^4\phi \exp \left\{- \lambda \left[\rho^2 - \left( 1 + \frac{\xi-1}{2\lambda} \right) \right]^2 - \sum_{m=1}^4 \xi \left(\phi_m - \frac{V_m}{\xi}\right)^2  \right\}  \, , \label{dp_phi_xi}
\end{equation}
where $\xi$ is an arbitrary parameter that will be determined to optimize the efficiency of the updating process.  The four real components $\phi_m$ are each generated according to
\begin{equation}
dp(\phi_m) \sim d\phi_m \exp\left[ -\xi \left( \phi_m - \frac{V_m}{\xi} \right)^2\right]  \label{dp_phim}
\end{equation}
and the total $\phi$ is accepted with a conditional probability
\begin{equation}
P = \exp\left\{ -\lambda \left[ \rho^2 - \left( 1 + \frac{\xi-1}{2\lambda} \right) \right]^2 \right\}  \, .  \label{conditional_probability}
\end{equation}
The parameter $\xi$ is now tuned to maximize the acceptance rate \cite{bunk}:
\begin{equation}
\xi^3 + (2\lambda-1)\xi^2 - 4\lambda\xi - 2\lambda\det(V) = 0  \, . 
\label{cubic_xi}
\end{equation}
In cases where $\lambda$ and $\det(V)$ are small, approximation schemes may be
used to solve Eq.~\eqref{cubic_xi}.  However, we want the Higgs updater to work
reliably (\textit{i.e.\ }with reasonable acceptance rates) for arbitrary $\kappa$
and $\lambda$.  Therefore, we calculate the exact positive solution for
Eq.~\eqref{cubic_xi} when the approximations break down.  Also, 
given $\lambda > 0$ and $\det(V) > 0$ there is only one positive solution for
Eq.~\eqref{cubic_xi}.

\begin{figure}[htb]
\centering
\includegraphics[scale=0.4]{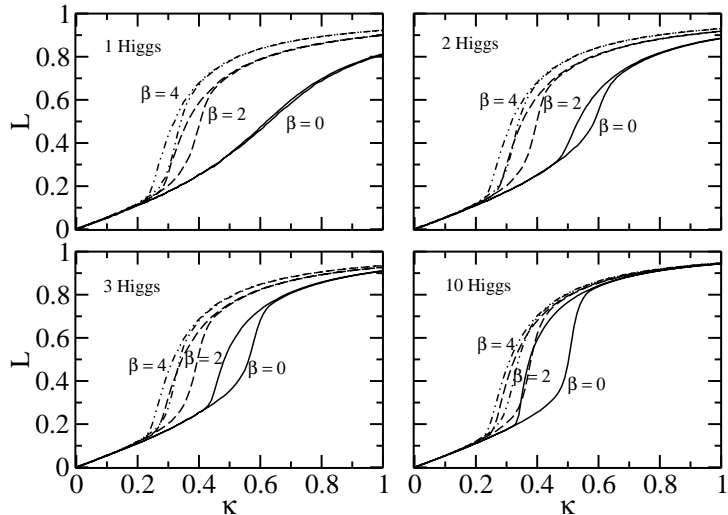}
\caption{Fast thermal cycles of the gauge-invariant link for $N=1,2,3,10$ Higgs fields on a $12^4$ lattice for $\lambda=1$ in the case where all Higgs fields have identical interactions and mass terms.  The solid curves represent $\beta=0$, dashed curves represent $\beta=2$, and dashed-dotted curves correspond to $\beta=4$. For the single-Higgs case with $\beta=0$ one can see that the hysteresis effect is small compared to the statistical fluctuations.}
\label{hyst_single}
\end{figure}

The phase structure of the $SU(2)$-Higgs theory can be probed by studying the way observables ({\it e.g.\ }the Higgs length, gauge-invariant link, and average plaquette) change with respect to the parameters $\beta$, $\kappa$, and $\lambda$.  Sudden changes in the behaviour (and even discontinuities in the case of first order phase transitions) of these quantities characterize the location and the strength of the phase transition.  While a variety of observables  could be used to diagnose the Higgs phase transition (dynamical terms in the action are frequently chosen), the gauge-invariant link is our preferred quantity for two main reasons.  First, it has a bounded range, between $-1$ and $1$, which sets a natural scale for comparing the strength of the phase transition across the entire parameter space of the theory.  Second, it is suitable in the extreme cases where $\lambda \rightarrow \infty$ (where the Higgs length goes to one), and $\beta \rightarrow \infty$ (where the gauge fields go to unity and the average plaquette becomes fixed).  Thus in contrast to the Higgs length and average plaquette,  the gauge-invariant link provides useful information on the Higgs phase transition for any $\beta$ and $\lambda$.

\begin{figure}[htb]
\centering
\includegraphics[scale=0.4]{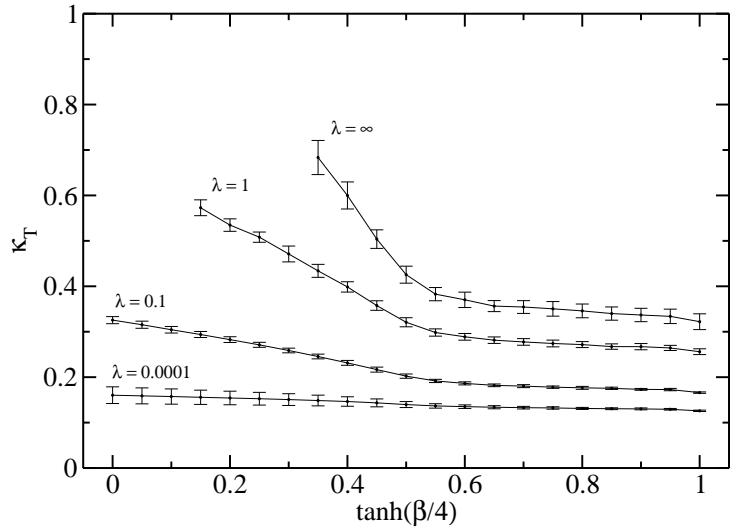}
\caption{Phase structure for the single-Higgs model on a $16^4$ lattice for selected values of $\lambda$.  The  parameter space above the phase boundary will be referred to as the Higgs region.
Note the termination of phase lines at small $\beta$ 
for sufficiently large $\lambda$, implying an analytic connection
between the Higgs and confinement phases.}
\label{single_phase_fig}
\end{figure}

A typical signature of a phase transition is a hysteresis curve resulting from a heating/cooling (thermal) cycle.  
Our study does not involve temperature, but we will use the phrase ``thermal cycle'' to mean the systematic increase, then decrease, of $\kappa$ as follows.
The gauge-Higgs coupling constant $\kappa$ can very loosely be thought of as an inverse temperature shared by the Higgs and gauge fields.  By iteratively increasing $\kappa$ after a number of Monte Carlo updates, where the field configurations for the previous $\kappa$ are used for the start of the new $\kappa$, and then iteratively decreasing $\kappa$, a hysteresis curve in an observable (such as the gauge-invariant link) indicates the presence of a phase transition.  The hysteresis curve can then be used to find the approximate location  $\kappa_T$ of the phase transition. This approach has  the advantage  that it does not require high statistics to extract information about the phase transition because it exploits the fact that the system does not thermalize at a phase transition with a small number of updates.
This allows efficient exploration of a large region of parameter space
for the $SU(2)$-Higgs model with a single Higgs \cite{lang,kuhnelt} and  is also well-suited to the multiple-Higgs case.

\begin{figure}[htb]
\centering
\includegraphics[scale=0.5]{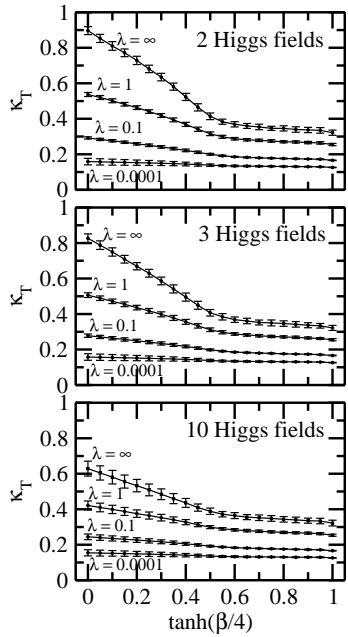}
\caption{Phase structure for the symmetric double-Higgs, triple-Higgs, and ten-Higgs models on a $16^4$ lattice for selected values of $\lambda$.
Note the continuation of phase lines down 
to $\beta=0$ in contrast to the single-Higgs model.}
\label{multiple_phase_fig}
\end{figure}

For a given $(\beta~,~\lambda)$, the location of the PT is initially located by  performing a ``fast'' thermal cycle for the gauge-invariant link,  increasing $\kappa$ from $0$ to $1$  and then decreasing back to $0$ in steps of $0.01$, using
one heatbath and one overrelaxation update for the gauge and Higgs fields at each value of
$\kappa$.\footnote{Approximately 10 updates are used to thermalize the  lattice  at the initial minimum value of $\kappa$ and re-thermalize it at the maximum value.}
A hysteresis curve resulting from the thermal cycle gives the approximate location of the phase transition $(\kappa_T \pm \sigma_{\kappa_T})_{fast}$  where $(\sigma_{\kappa_T})_{fast}$ represents the region where the difference in the gauge-invariant link between the heating/cooling runs is greater than the estimated statistical fluctuations. The value of $\kappa_T$ is then refined by narrowing in on the region around the PT,  $\kappa \in [(\kappa_T-2\sigma_{\kappa_T})_{fast},(\kappa_T+2\sigma_{\kappa_T})_{fast}]$, and performing a slow heating cycle with the same number of updates and steps in $\kappa$ as before, but with correspondingly  smaller intervals.  The refined location of the PT and its corresponding uncertainty is then $(\kappa_T \pm \sigma_{\kappa_T})_{slow}$.
We emphasize that the signal of a  PT is a hysteresis gap that exceeds the estimated statistical fluctuations within the thermal cycles.

\begin{figure}[htb]
\centering
\includegraphics[scale=0.35]{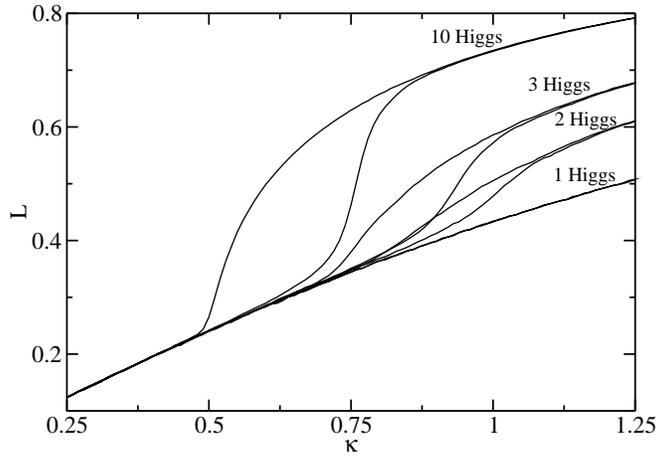}
\caption{Fast  thermal cycles of the gauge-invariant link for $N=1,2,3,10$ Higgs fields on a $16^4$ lattice for $(\beta=0,\lambda=\infty)$,  in the case where all Higgs fields have identical interactions and mass terms. 
For the single-Higgs case with $\beta=0$ one can see that the hysteresis effect is small compared to the statistical fluctuations.
  }
\label{hyst_strength}
\end{figure}

\bigskip

\begin{figure}[htb]
\centering
\includegraphics[scale=0.4]{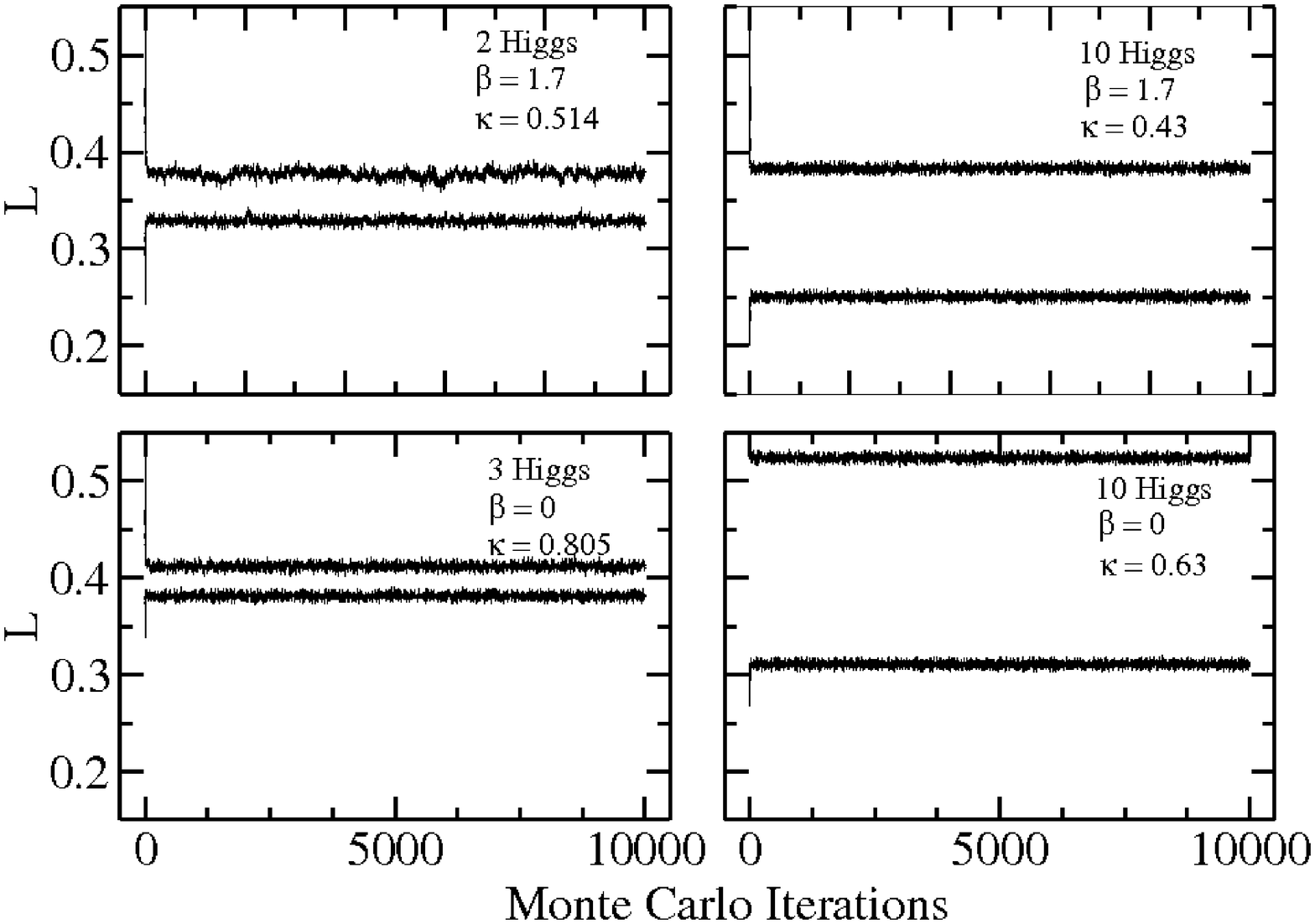}
\caption{Metastability of the gauge-invariant link at selected transition points for $\lambda=\infty$ on a $16^4$ lattice. }
\label{metastability_fig}
\end{figure}

To validate our methodology we first consider the single-Higgs case.  The results for (fast) thermal cycles
of the gauge-invariant link on a $12^4$ lattice for $\lambda=1$ and selected values of $\beta$ are shown in Figure~\ref{hyst_single}.
Figure~\ref{single_phase_fig}  presents  our results for the phase structure for the single-Higgs case on a $16^4$ lattice.  The termination of the phase line for $\lambda=1$ is associated with the absence of a hysteresis curve for $\beta=0$ and the presence of one for $\beta=2$ in Fig.~\ref{hyst_single}. For later reference, we will refer to the region of parameter space above the phase boundary as the Higgs region.  Figure~\ref{single_phase_fig} compares favourably with the Ref.~\cite{kuhnelt} results on smaller lattices, and the analytic connection that exists between the phases is expected from Ref.~\cite{fradkin}.
In addition, we have reproduced (not shown) the  bimodal distribution functions of Ref.~\cite{bock} for the  gauge-invariant  link used as a  criterion for  studying PTs  \cite{languth,languth2,bock,campos}.

\begin{figure}[htb]
\centering
\includegraphics[scale=0.5]{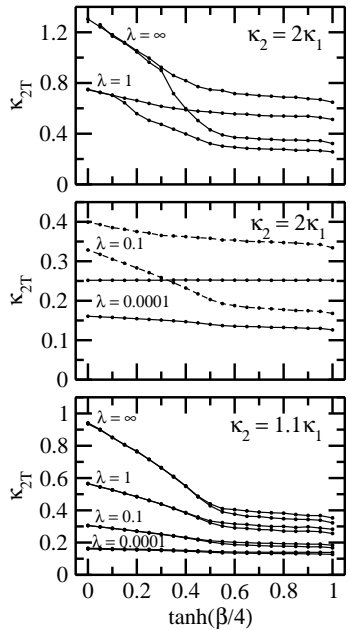}
\caption{Phase structure for asymmetric two-Higgs models on a $12^4$ lattice for selected values of $\lambda$. Error bars associated with determination of the transition points have been suppressed for clarity.  The lower curve in each pair represents the phase line extracted from the gauge-invariant link for the Higgs field with the larger $\kappa$; the upper phase line is associated with the Higgs field with smaller $\kappa$.  }
\label{asymmetric_phase1}
\end{figure}

We now consider the symmetric multi-Higgs case of $N$ Higgs fields where  $\kappa_n=\kappa$ for all $n$ 
({\it i.e.\ }the discretized theory has a global permutation symmetry, $S_N$).
As illustrated in Figure~\ref{hyst_single}, when $N\ge 2$ hysteresis effects exist even for $(\beta=0,\lambda=1)$.\footnote{For $N$ Higgs fields in the symmetric case, we calculate the average of the gauge-invariant link over all fields.  We have verified that the expectation value of the gauge-invariant link is identical for all fields as expected from the $S_N$ permutation  symmetry of the discretized theory.} This implies a completion of the phase boundaries for the multi-Higgs case as shown in 
Figure~\ref{multiple_phase_fig}, in contrast to the single-Higgs case where the Higgs and confinement regions maintain an analytic connection.  
The volume dependence of this conclusion  has been explored by comparing the phase structure for $12^4$, $16^4$ and $32^4$ lattices; there would be no discernable difference between these lattices within Figure~\ref{multiple_phase_fig}.

\begin{figure}[htb]
\centering
\includegraphics[scale=0.4]{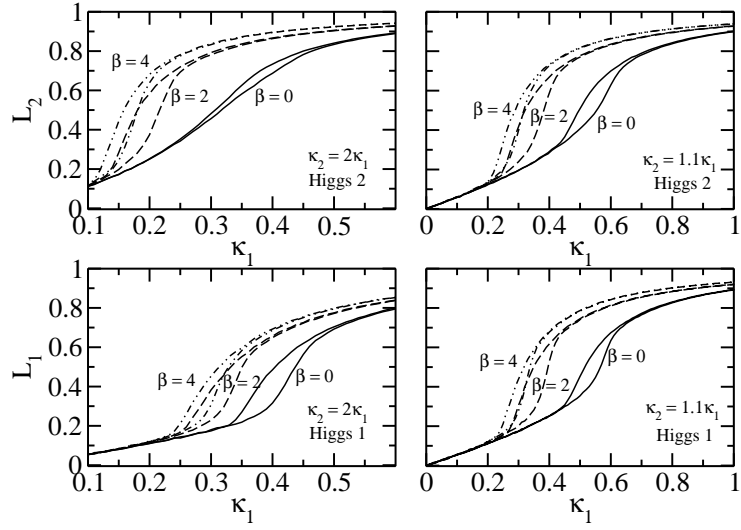}
\caption{Fast thermal cycles of the gauge-invariant link for each Higgs field  in  asymmetric two-Higgs models
for $\lambda=1$ and selected values of $\beta$ on a $12^4$ lattice. Solid curves correspond to $\beta=0$, dashed curves correspond to $\beta=2$ and dashed-dotted curves correspond to $\beta=4$. }
\label{hyst_asym}
\end{figure}

The dependence of the PT strength on $N$ can be assessed qualitatively from the $(\beta=0,\lambda=\infty)$  thermal cycles on a $16^4$ lattice shown in Fig.~\ref{hyst_strength}.  The  hysteresis curve commences for  $N=2$ and becomes more  pronounced as $N$ increases, indicative of increasing
strength of the PT with increasing $N$.\footnote{The same conclusion can be extracted from Fig.~\ref{hyst_single} which  also shows that the PT strength is largely unaffected for larger $\beta$.}   Figure~\ref{metastability_fig}
demonstrates that the signal of a first-order PT, which is difficult to ascertain in the single-Higgs case \cite{jersak,languth,languth2,bock,campos,tomiya,bonati},
becomes increasingly evident with increasing $N$ in the $\lambda=\infty$ metastability curves for the gauge-invariant link obtained from hot/cold starts of the system at the transition point ($\kappa=\kappa_T$).

Thus the PT of the symmetric multi-Higgs $SU(2)$-Higgs model is enhanced for small $\beta$ as additional Higgs fields are incorporated into the model.   To explore the  possibility that this enhancement originates from overlapping phase transitions as found in the three-dimensional $U(1)$ model \cite{ono}, we consider an asymmetric two-Higgs case where $\kappa_1\ne\kappa_2$.  Figure~\ref{asymmetric_phase1} shows the  phase structure  extracted  from the
gauge-invariant link for each Higgs field in the case of a large asymmetry $\kappa_2=2\kappa_1$.  There is an alignment of the PTs for $\lambda=1$ and $\lambda=\infty$ in the $\beta$ region  where the phase lines terminate in the single-Higgs model.   Furthermore,  for $\lambda=1$ and $\lambda=\infty$ in the non-aligned region  the lower phase line (corresponding to the Higgs field with  larger $\kappa$) closely resembles the single-Higgs case, while the upper phase line (corresponding to the Higgs field with smaller $\kappa$) is similar to the large $\beta$ single-Higgs model. This behaviour suggests that the first field to 
enter the Higgs region does so similar to a single-Higgs system, and influences the PT of the second field.
Figure~\ref{asymmetric_phase1} also shows the  phase structure  extracted  from the
gauge-invariant link for each Higgs field in the case of a smaller asymmetry $\kappa_2=1.1\kappa_1$.  At this level of asymmetry the phase lines are much more closely aligned, and the phase structure is virtually indistinguishable from the symmetric two-Higgs case.  It should be noted that the full theory would experience multiple PTs in the non-aligned regions of parameter space.

\begin{figure}[htb]
\centering
\includegraphics[scale=0.4]{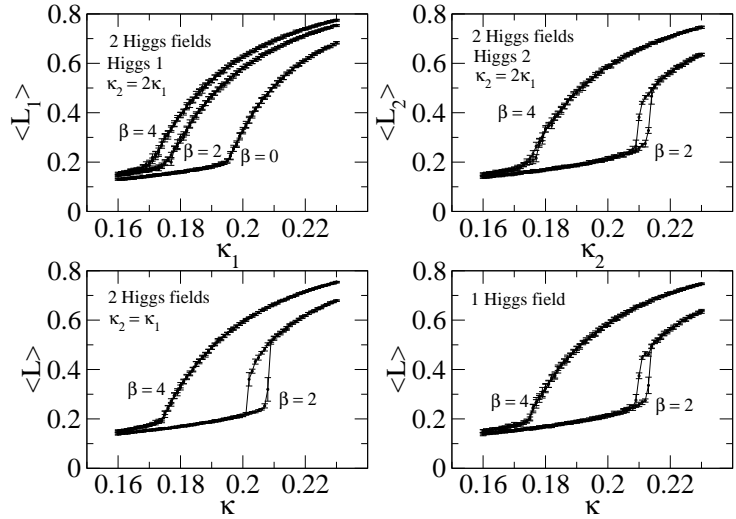}
\caption{Slow thermal cycles on an $8^4$ lattice for the gauge-invariant link(s) for the large asymmetry ($\kappa_2=2\kappa_1$) two-Higgs model, the symmetric two-Higgs model, and the single-Higgs model for $\lambda=0.1$ and selected values of $\beta$.}
\label{new_fig}
\end{figure}

The strength of the PTs in the asymmetric two-Higgs model can  be examined through the $\lambda=1$  hysteresis curves in Figure~\ref{hyst_asym} in  comparison with Figures~\ref{hyst_single} and \ref{hyst_strength}. In the small asymmetry case, it is evident that the PTs  for each Higgs field are comparable in strength to the symmetric two-Higgs case.  For a large asymmetry,  the first (larger $\kappa$) Higgs field 
to enter the Higgs region
resembles  the  single-Higgs system with a slight enhancement of strength, although this enhancement still results in a weaker PT than the symmetric two-Higgs case.  
However, the  PT of the second Higgs field entering the Higgs region in the large asymmetry case behaves like a cold (large $\beta$) single Higgs system.\footnote{In making this comparison it is important to take into account the different scales for $\kappa$ in Figs.~\ref{hyst_single} and \ref{hyst_asym}.}  
This point is further illustrated in Figure~\ref{new_fig} where very slow thermal cycles for the gauge-invariant link  are performed for $\lambda=0.1$  for the two different Higgs in the large asymmetry case, the 
two-Higgs symmetric case and the single Higgs system.  The very slow thermal cycles were performed with 10 updates (one heatbath and one overrelaxation) to thermalize the lattice, 20 updates which are used to calculate expectation values of the gauge-invariant link, and (small) steps of $0.001$ in $\kappa$.  In particular, the
$\beta$ dependence of the gauge-invariant link is greatly diminished for the second Higgs entering the Higgs region, and for $\beta=2$ the first-order PT signal  is absent for the second Higgs.  This behaviour can be understood as a cooling of the gauge fields as a result of the PT of the first Higgs entering the Higgs region.  
Based on  the comparative behaviour of the large- and small-asymmetry two-Higgs systems, we thus conclude that as the two-Higgs system approaches the symmetric limit, there will exist a mutual reinforcement of the PT strength.

In summary, we have studied the effect of multiple Higgs fields on the phase structure of the $SU(2)$-Higgs model.
The striking qualitative observation is that the analytic connection between the Higgs and confinement regions in the single-Higgs model is not present in the $N\ge 2$ Higgs models used in this work.  Since our study is restricted to the action of Eq.~\eqref{lattice_S}, where all individual terms containing more than one Higgs flavour are absent, it is possible that the addition of those extra terms will reveal an analytic connection between the Higgs and confinement regions elsewhere in that larger parameter space.  That future exploration will allow for an interesting discussion of symmetry breaking.
Furthermore, we see evidence for a progressive increase in the strength of the PT for increasing numbers of Higgs fields,  and progressively stronger signals of first-order PTs as $N$ increases.
Two-Higgs models with asymmetric hopping parameters
reveal that this behaviour can be attributed to a reinforcement from overlapping  PTs
 as found in  three-dimensional multi-Higgs $U(1)$ models \cite{ono}.  
Remarkably, a moderate 10\% asymmetry in the hopping parameters does not prevent this enhancement, indicating that fine-tuning of multi-Higgs models is not needed to realize the enhancement effects of overlapping phase transitions.

We are grateful for research support from the Natural Sciences and Engineering Research Council of Canada (NSERC).
This work was made possible by the facilities of WestGrid (www.westgrid.ca) and the Shared Hierarchical Academic Research Computing Network (SHARCNET:www.sharcnet.ca). TGS thanks Kurt Langfeld for discussions regarding the results of Ref.~\cite{fradkin}. RL thanks Kei-ichi Nagai for conversations.  We appreciate the valuable critical reading by R.M.~Woloshyn of a draft of this manuscript.


\clearpage

\begin{thebibliography}{99}

\bibitem{gunion} J.F.~Gunion, H.E.~Haber, Nucl.~Phys.~B272 (1986) 1; Erratum-ibid.B402 (1993) 567;\\
J.F.~Gunion, H.E.~Haber, G.L.~Kane, S.~Dawson, ``The Higgs Hunter's Guide'' (Addision-Wesley, 1990).


\bibitem{fradkin}
E.~Fradkin, S.H.~Shenker, Phys.~Rev.~D19 (1979) 3682;\\
K.~Osterwalder, E.~Seiler, Ann.~Phys.~110 (1978) 440.

\bibitem{lang} C.B.~Lang, C.~Rebbi, M.~Virasoro, Phys.~Lett.~104B (1981) 294.

\bibitem{kuhnelt} H.~K\"uhnelt, C.B.~Lang, G.~Vones, Nucl.~Phys.~B230 (1984) 16.

\bibitem{elitzur}S.~Elitzur, Phys.~Rev.~D12 (1975) 3978.


\bibitem{greensite} J.~Greensite, \v{S},~Olejn\'ik, Phys.~Rev.~D74 (2006) 014502;\\
W.~Caudy, J.~Greensite, Phys.~Rev.~D78 (2008) 025018.



\bibitem{jersak} J.~Jers\'ak, C.B.~Lang, T.~Neuhaus, G.~Vones, Phys.~Rev.~D32 (1985) 2761.

\bibitem{languth} W.~Langguth, I.~Montvay, Phys.~Lett.~165B (1985) 135.

\bibitem{languth2} W.~Langguth, I.~Montvay, P.~Weisz, Nucl.~Phys.~ B277 (1986) 11.

\bibitem{bock} W.~Bock, H.G.~Evertz, J.~Jers\'ak, D.P.~Landau, T.~Neuhaus, J.L.~Xu, Phys.~Rev.~D41 (1990) 2573.


\bibitem{campos} I.~Campos, Nucl.~Phys.~B514 (1998) 336.

\bibitem{tomiya} M.~Tomiya, T.~Hattori, Phys.~Lett.~140B (1984) 370.


\bibitem{bonati} C.~Bonati, G.~Cossu, A.~D'Alessandro, M.~D'Elia, A.~Di Giacomo, arXiv:0901.4429 [hep-lat].


\bibitem{ono} M.N.~Chernodub, E.-M.~Ilgenfritz, A.~Schiller,  Phys.\ Rev.\ B73, 100506 (2006);\\
M.~Bock, M.N.~Chernodub, E.-M.~Ilgenfritz, A.~Schiller, Phys.\ Rev.\ B76, 184502 (2007);\\
T.~Ono, I.~Ichinose, T.~Matsui, Proc.~Sci.~LATTICE2008 (2008) 252.



\bibitem{heatbath} M.~Creutz, Phys.~Rev.~D21 (1980) 2308;\\
A.D.~Kennedy, B.J.~Pendleton,  Phys.~Lett.~156B (1985) 393.


\bibitem{bunk} B.~Bunk, Nucl.~Phys.~B (Proc.~Suppl.) 42 (1995) 566.


\bibitem{fodor} Z.~Fodor, J.~Hein, K.~Janzen, A.~Jaster, I.~Montvay, Nucl.~Phys.~B439 (1995) 147.

\bibitem{over_relax} M.~Creutz, Phys.~Rev.~D36 (1987) 515;\\
Z.~Fodor, K.~Jansen, Phys.~Lett.~B331 (1994) 119;\\
K.~Kajantie, M.~Laine, K.~Rummukainen, M.E.~Shaposhnikov, Nucl.~Phys.~B466 (1996) 189.





\end{thebibliography}
\end{document}